\documentclass[12pt]{article}
\usepackage[utf8]{inputenc}
\usepackage{fullpage}
\usepackage{graphicx}
\usepackage{epstopdf}
\usepackage{multirow}
\usepackage{authblk}
\usepackage{hyperref}

\title{Growth in the number of references in engineering journal papers during the 1972-2013 period}

\author[1]{Iñaki Ucar}
\author[2]{Felipe López-Fernandino}
\author[2]{Pablo Rodriguez-Ulibarri}
\author[2]{Laura Sesma-Sanchez}
\author[2]{Veronica Urrea-Micó}
\author[2]{Joaquín Sevilla \thanks{joaquin.sevilla@unavarra.es}}

\affil[1]{\small Dept. of Automatics and Computer Science}
\affil[2]{\small Dept. of Electrical and Electronic Engineering}
\affil[ ]{\small Public University of Navarre}

\date{
	\small Submitted to Springer's Scientometrics on June 18, 2013\\
	Accepted on August 15, 2013 | Published online on August 31, 2013\\
	DOI: \href{http://dx.doi.org/10.1007/s11192-013-1113-6}{10.1007/s11192-013-1113-6}
}

\newcommand{\keywords}[1]{\par\vspace{3mm}\noindent{\small{\bf Keywords\/}: #1}}

\begin{document}

\maketitle

\begin{abstract}
\noindent The number of references per paper, perhaps the best single index of a journal's scholarliness, has been studied in different disciplines and periods. In this paper we present a four decade study of eight engineering journals. A data set of over 70000 references was generated after automatic data gathering and manual inspection for errors. Results show a significant increase in the number of references per paper, the average rises from 8 in 1972 to 25 in 2013. This growth presents an acceleration around the year 2000, consistent with a much easier access to search engines and documents produced by the generalization of the Internet.

\end{abstract}

\section{Introduction}\label{intro}

Along the 20th century, scientific communication has become highly structured. Journal papers, the main tool of this communication, are written in the ``IMRAD'' structure \cite{day1979write}. In this structure, references play a very important role. In order to keep the papers short, the information included must be restricted to the novelty of the presented research, while all the contextual information needed to understand this novelty is just referenced.

\pagebreak\noindent It has been said that the ``density of citation'' (i.e., the number of references per paper) is perhaps the best single index of a journal's scholarliness, in the sense that its authors rely more or less heavily upon their knowledge of the work of others \cite{xhignesse1967bibliographical}. De Solla Price \cite{price1970citation} suggested the idea that the sheer amount of bibliographical references can provide ``a handy measure of social linkage''. However, little attention has been paid to the study of the number of references \cite{raeyb}. The purpose of this work is to make a contribution to this knowledge by studying the time evolution of the number of references per paper over a four decade period in engineering journals.

Previous studies have been made for different disciplines over different periods of time, showing in general significant increases in the number of references that papers include. In the extensive work by Yitzhaki and Ben-Tamar \cite{raeyb}, the Journal of Biological Chemistry is analyzed from 1910 to 1985. They performed paper sampling of regular research papers for certain years. The sample size varied from 40 to 60 papers out of earlier volumes (1910, 1920), up to several hundreds out of later volumes (1930-1985), whose populations of papers were much greater. The investigation reveals that all the subject fields checked show a steady increase in the number of references per paper with time, particularly in the last two decades.

Lin and Huang \cite{raeyc} analyze the number of references in environmental engineering articles by year, in a 10-year interval (1999-2008) and demonstrate a slightly increasing tendency, with an average number of 33.08 references.

Peritz and Bar-Ilan \cite{raeyd} carried out a reference analysis of the journal Scientometrics in order to investigate the interdisciplinarity of bibliometrics-scientometrics. They concluded that the field relies heavily on itself, on library and information science and on sociology, history and philosophy of science. The study is based on 169 papers published between 1990 and 2000. The mean number of references in 1990 was 15.1, and in 2000 it increased to 19.8. They state that, for more conclusive results, studies of longer periods of time should be conducted.

With the 25th volume of Scientometrics publication, Wouters and Leydesdorff \cite{wouters1994hasprice} made a combined bibliometric and social network analysis about this journal. They examined the period from 1978 to 1992, determining a linear increase in the number of articles per year, as well as the number of references per year. Nevertheless, the number of references per article remained stable at an average of 15 references from 1985 onward without any meaningful variation. Papers related to Psychology were studied by Krampen \cite{raeyf} presenting a significant increase in the number of references in a 20-year period sampled in three particular years (1985, 1995 and 2005).

This quick glance at the work in the field clearly shows that the number of references per paper increases over time almost in all disciplines and periods of time. There are a number of possible reasons for this general increase, as has already been pointed out by previous authors. In this paper, we aim to quantify this time increment particularly for engineering journals and also to evaluate the possible influence of the advent of the Internet.

Over the last twenty years, the volume of information available for the general public has drastically increased. Specifically, scientific researchers have found major benefits from this new information channel. The Internet allows them not only to track easily the latest advances in their research topic, but also to spread their contributions worldwide. Therefore, the hypothesis formulated here is that, due to the huge amount of information available nowadays, a recent paper has more chances to include more references than older ones.

\section{Methodology}\label{sec:1}

In order to test the possible effect of using the Internet on the amount of references per paper, we have to analyze a time span wide enough, and having the Internet birth approximately in the center. The selected period ranges from 1972 to 2013. Other similar works cover shorter time ranges \cite{raeyc}, \cite{ASI:ASI21448} or, in the case of wide time periods, they proceed to sample the years of study \cite{raeyb}, \cite{raeyf}. In our case, all the years of the span are included.

The study here presented intends to offer a general view of engineering, therefore the number of journals considered had to be enough to get the big picture. On the other hand, to keep the size of the data set manageable, especially considering that the time span extends over 40 years, the number of journals had to be limited. A compromise value of 8 journals was established. The journals were selected among high-impact and long-standing IEEE journals of various fields. The final selection is shown in Table~1.

\begin{table}[ht]\label{tab:1}
	\footnotesize \centering
	\caption{Selected journals.}
	\begin{tabular}{llc}
		\hline\noalign{\smallskip}
		\bfseries Title & \bfseries Abbrev. & \bfseries Years \\
		\noalign{\smallskip}\hline\hline\noalign{\smallskip}
		IEEE Transactions on Antennas and Propagation & ITAP & 1963- \\
		IEEE Transactions on Biomedical Engineering & ITBE & 1964- \\
		IEEE Transactions on Broadcasting & ITB & 1963- \\
		IEEE Transactions on Communications & ITComm & 1972- \\
		IEEE Transactions on Computers & ITComp & 1968- \\
		IEEE Transactions on Information Theory & ITIT & 1963- \\
		IEEE Transactions on Instrumentation and Measurement & ITIM & 1963- \\
		IEEE Transactions on Microwave Theory and Techniques & ITMTT & 1963- \\
		\noalign{\smallskip}\hline
	\end{tabular}
\end{table}

\noindent Data collection was performed by running an ad-hoc homemade Python script. This script is able to conduct searches on the ISI Web of Knowledge (WoK) search service and automatically export a full record of all the results, 500 by 500 (the limit imposed by WoK). The data set was retrieved on the 10th of April 2013 and comprises 151 files with all papers from the selected journals stored by the Inspec database. These files were postprocessed in order to extract the number of references and the number of pages per paper since 1972.

As previously reported \cite{raeyc}, these automatically generated data sets may contain a significant number of errors, particularly in the older files that have been generated with OCR systems fed with the paper editions of the journals. In order to check possible mistakes, a random set of papers with a too low number of references (0, 1 and 2) were inspected by hand. This process showed that some of these papers are comments or erratum, a few has a wrong number of references and some others are valid research articles. On the other side, all the very high values (more than 150 references) were also inspected manually and two of them were corrected (one paper with 3838 references which were actually 38 and another with 639 which were 39). These high values are mainly review articles and bibliographic compendiums; in any case, they only represent about $0.06$ \% of the population. We are primarily interested in research articles; however it was decided not to exclude any item using this criterion. Although papers with higher number of references tend to be reviews, and those with lower ones erratum or comments, there is no way of deciding the kind of document based solely on the number of references \cite{de1965networks}, and a manual classification of all the records would be too time consuming. It is assumed that the distribution of the different kind of papers is sufficiently constant over time, although further work would be required.

It has been found, however, that the number of pages field contains much more errors. Accordingly, papers with less than one page were removed. In a similar way, papers with an absurdly large number of pages were filtered using the outlier threshold proposed by \cite{laurikkala2000informal}:

\begin{equation}
[Q_1-k(Q_3-Q_1), Q_3+k(Q_3-Q_1)]
\end{equation}

\noindent where $Q_1$ and $Q_3$ are the lower and upper quartiles respectively, and the constant was fixed at a permissive value of $k=3$.

The final data set comprises 70036 articles of these 8 journals on a 40-year interval. The details are listed in Table~2. All the data collected and scripts made for this work are available upon request.

\begin{table}\label{tab:2}
	\footnotesize \centering
	\caption{Data set.}
	\begin{tabular}{lrrrrr}
		\hline\noalign{\smallskip}
		\multirow{4}{*}{\bfseries Journal} & \multicolumn{5}{c}{\bfseries Articles} \\
		\noalign{\smallskip}\cline{2-6}\noalign{\smallskip}
		& \multirow{2}{*}{\bfseries Retrieved} & \multicolumn{3}{c}{\bfseries Removed} & \multirow{2}{*}{\bfseries Total} \\
		\noalign{\smallskip}\cline{3-5}\noalign{\smallskip}
		& (all years) & before 1972 & $<1$ pages & outliers & (1972-2013) \\
		\noalign{\smallskip}\hline\hline\noalign{\smallskip}
		ITAP & 12441 & 133 & 5 & 7 & 12296  \\
		ITBE & 7432 & 63 & 3 & 3 & 7363 \\
		ITB & 1627 & 6 & 0 & 12 & 1609 \\
		ITComm & 10792 & 0 & 9 & 11 & 10772 \\
		ITComp & 6716 & 202 & 6 & 41 & 6467 \\
		ITIT & 10427 & 161 & 5 & 341 & 9920 \\
		ITIM & 8593 & 25 & 2 & 3 & 8563 \\
		ITMTT & 13281 & 217 & 5 & 13 & 13046 \\
		\noalign{\smallskip}\hline\noalign{\smallskip}
		\bfseries Total & 71309 & 807 & 35 & 431 & 70036 \\
		\noalign{\smallskip}\hline
	\end{tabular}
\end{table}
       
\section{Results \& Discussion}\label{sec:2}

The number of references in all the collected papers shows a monotonic increase over time, as can be seen in Figure~1, where all collected data are plotted. In order to deal with over-plotting and improve data representation, every point in the mentioned figure was jittered in both axes adding a uniformly distributed component within $\pm 0.5$ units. The yearly average number of references per paper is presented in Figure~1 as a white line superimposed to the individual data representation.

\begin{figure}[t]
  \centering
  \includegraphics[width=\textwidth]{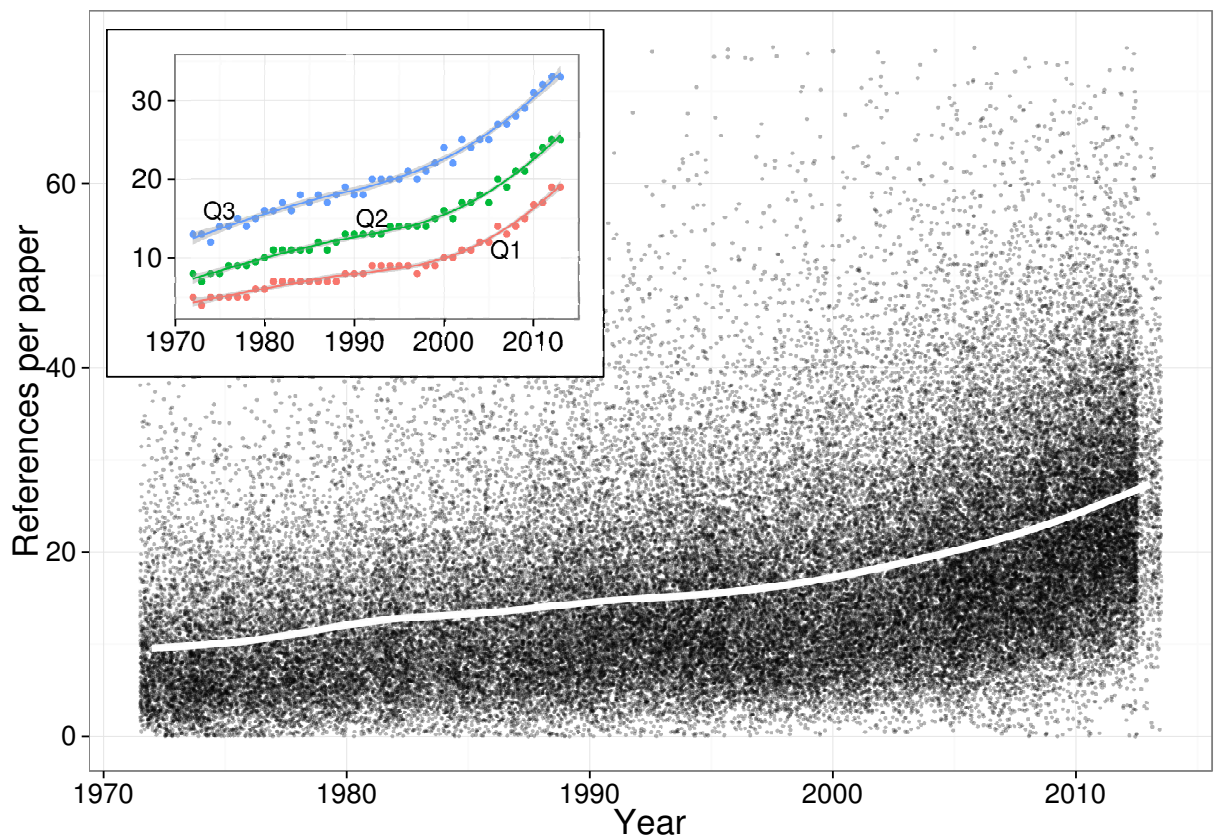}
  \caption{Time evolution of the number of references per paper, data aggregated from the 8 journals studied. In order to get a grey scale proportional to the actual number of papers, overprinting is avoided adding a random jitter. The white line is the yearly average, while the insert shows three lines representing quartiles of the distribution.}
  \label{fig:1}
\end{figure}

Moreover, a trend change can be seen around the year 2000. The inner plot of Figure~1 emphasizes this turning point. The three lines represent the three quartiles of the data distribution, and it can be seen that they clearly undergo a slope change. Overall, the median grows from 8 references per paper in 1972 to 16 in 2000, and from there to 25 references per paper in 2013.

Figure~2 presents a scatter plot for each of the eight studied journals separately. The representation is made in the same way as that of Figure~1: with jitter in the data plotting and a white line representing the yearly average number of references. All of them exhibit a sustained growth. However, the ``year 2000 effect'' is not present in all cases, showing different shapes for different journals. Those of \textit{Broadcasting} and \textit{Instrumentation and Measurement} are particularly significant.

\begin{figure}[ht]
  \centering
  \includegraphics[width=\textwidth]{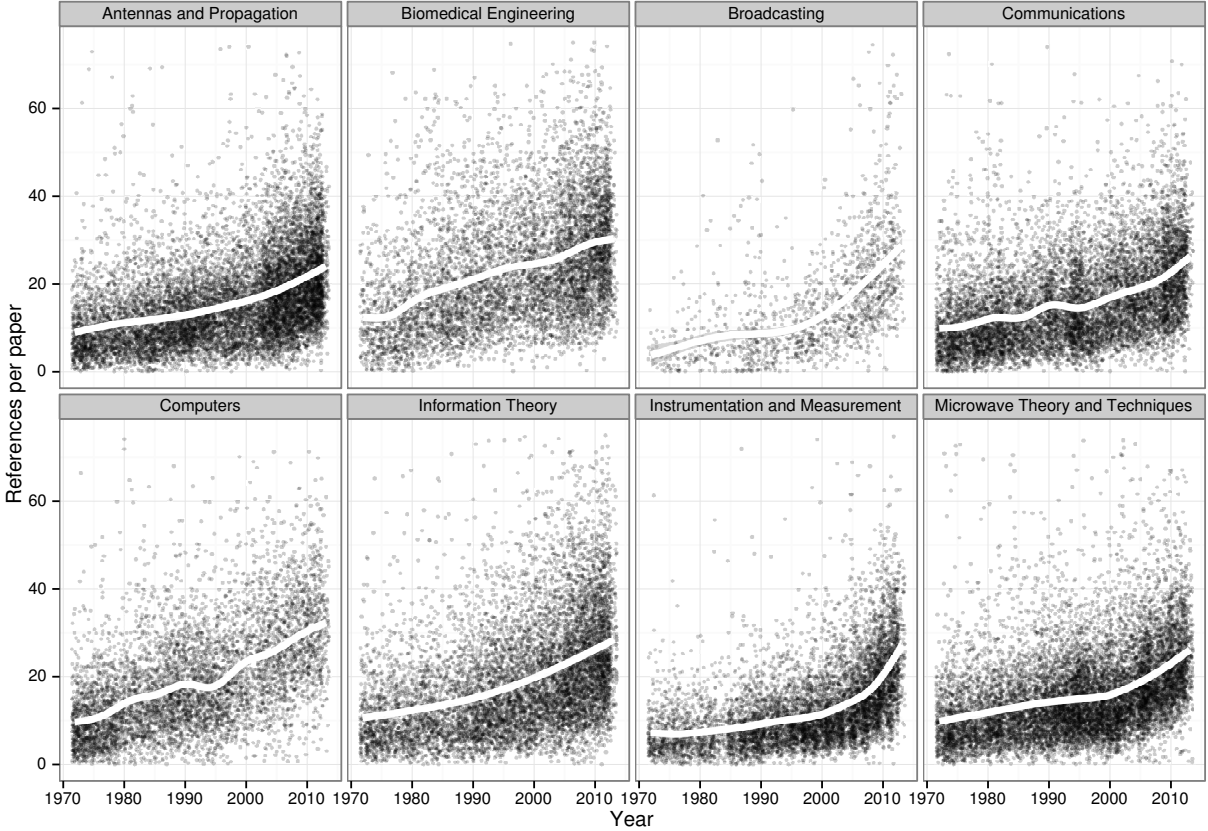}
  \caption{Time evolution of the number of references per paper. Data from each studied journal are presented here in separate plots.}
  \label{fig:2}
\end{figure}

The yearly distribution of the number of references per paper is presented in Figure~3 in two ways: the probability density function (PDF), on the left, and the cumulative density function (CDF), on the right. The CDF readily demonstrates the aforementioned rise in the average number of references per paper. Meanwhile, the PDF shows a widening as the average grows.

The distribution of the number of references is discrete, which suggests that it belongs to the Poisson family. It has been found a high index of overdispersion in all years, with a mean variance-to-mean ratio (VMR) of $8.64$. In fact, per-year distributions, as well as overall distribution, closely follow a Negative Binomial (NB), a good candidate for overdispersed Poisson processes. Unfortunately, this NB fit is not statistically significant in any case, which implies that another Poisson mixture model with 3 or more free parameters shall be best suited to describe this population.

\begin{figure}[ht]
  \centering
  \includegraphics[width=\textwidth]{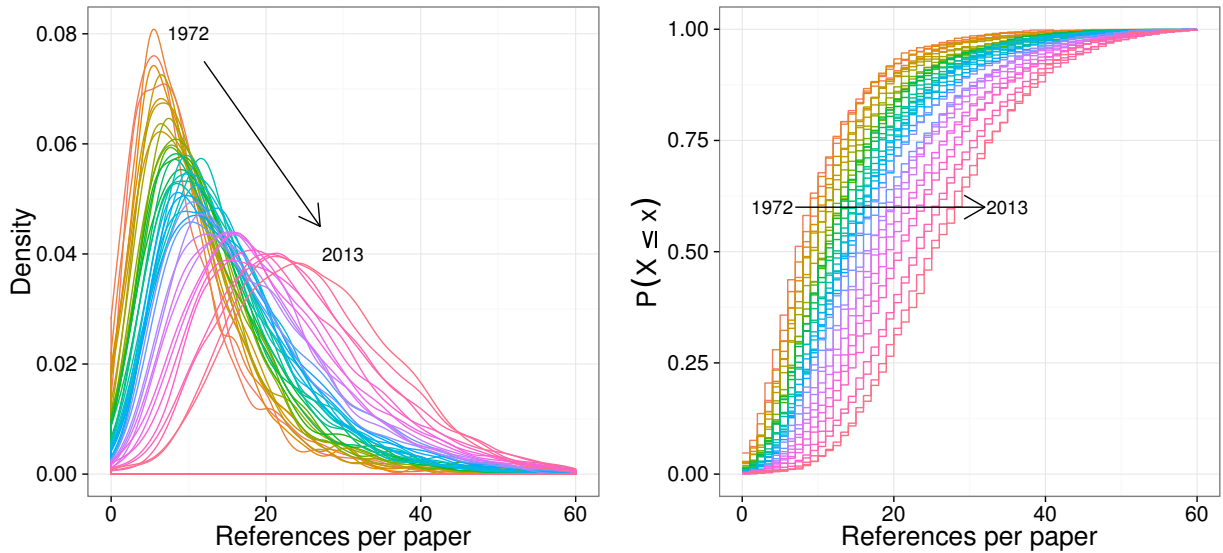}
  \caption{Yearly distribution of the number of references per paper: a) probability density function (PDF), and b) cumulative density function (CDF).}
  \label{fig:3}
\end{figure}

One of the reasons that could account for the increase in the number of references that the papers include could be their length increment \cite{raeyb}. As the data set we collected also included the number of pages of each document, this possible correlation was studied.

When plotting references against pages (see insert of Figure~4), a clear correlation can be seen, although it becomes weaker as the number of pages exceeds 10. The slope of the linear fitting of this correlation gives a value of references per page that can be calculated for the papers of each year. These data are presented in Figure~4. This measure of density of citation per page presents a significant growth over time, from around 1.75 in 1972 to 2.50 nowadays. Superimposed to this long range growth, there are short range fluctuations that make difficult to assign unambiguously a functional form to the data set. However, the year 2000 effect can be seen in this representation. This fact allows us to state that, although the growth in the number of pages probably contributes to the references increase, an independent effect (or effects) causes an accelerated growth in the number of references over time.

\begin{figure}[ht]
  \centering
  \includegraphics[width=\textwidth]{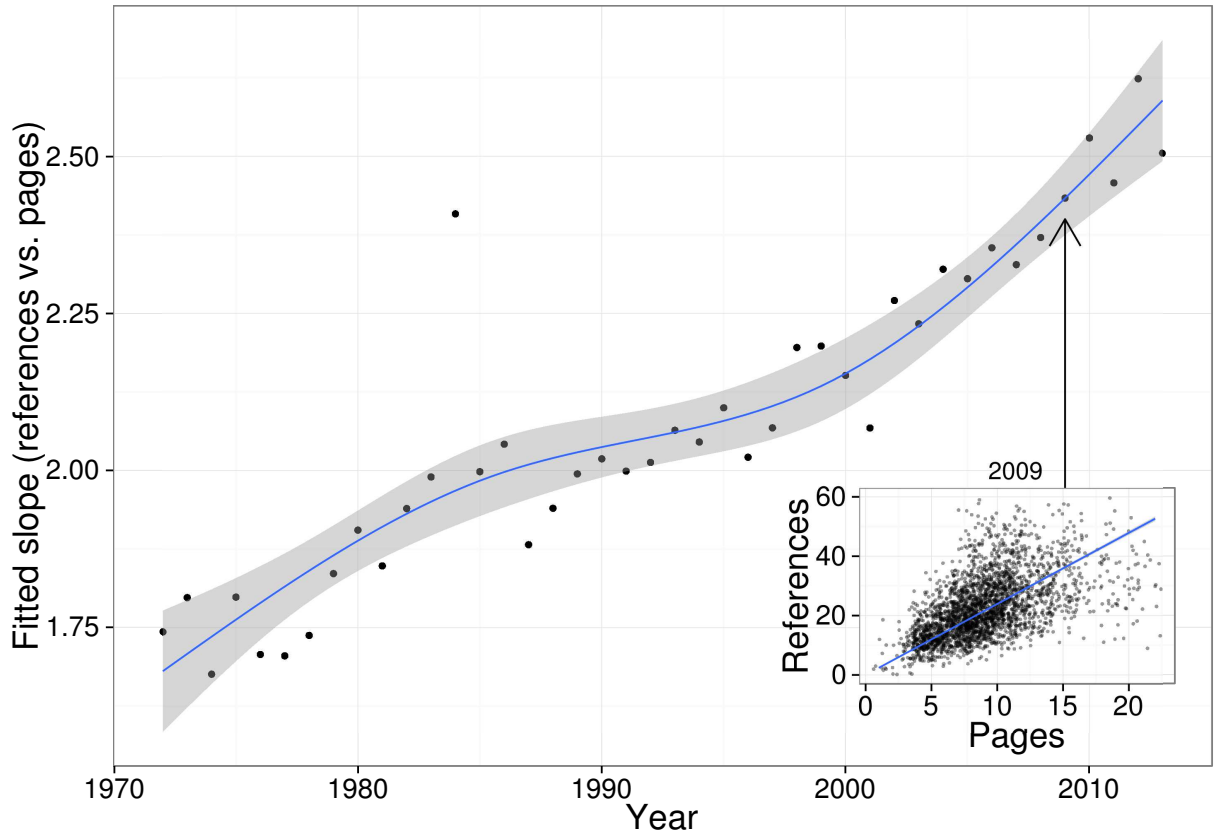}
  \caption{Time evolution of the number of references per page. Each dot stands for the slope of the linear fitting of references against pages of all data corresponding to a particular year; the insert shows the 2009 case. The line is generated as a 3rd order polynomial approximation, with standard error shown as a shadowed region.}
  \label{fig:4}
\end{figure}

There are many factors affecting the number of references in a paper that have already been identified by different authors. We have already discussed the type and length of the documents. Let us confront our results with the rest of those factors.

The intrinsic nature of the field \cite{raeyb} is relevant; the tradition in different fields yields to distinct citation habits. The eight journals selected cover fields from antennas and propagation to computers, which cannot be considered as the same field. However, all of them lie in a broad category of engineering, in fact all the journals are branches of the same organization (IEEE), where the citation habits can be considered sufficiently homogeneous. However, further study is needed to prove this assumption.

The cumulative characteristic of scientific knowledge generates an increasing body of information and, consequently of citable literature. Although this has been identified as a reason for the increase in the number of references per paper over time \cite{raeyb}, it is not obvious why each particular novelty, each paper, would need a wider base of knowledge to be supported.

\noindent Persson and coworkers \cite{persson2004inflationary} have studied the effect of collaboration. They noted that the productivity (papers per author) is growing faster than the number of publications; therefore, an intensifying scientific collaboration and an increasing density of co-publication networks is the only possible explanation. It is reasonable to assume that adding authors to papers also means adding references to papers, since authors can be assumed to have different exposure and access to literature. As the number of coauthors was not in our original data set, we have not checked this possible correlation.

Shearer and Moravcsik \cite{shearerCitation1979} suggest that over the recent decades it has become fashionable and perhaps even politically advantageous to cite more references. ``Courtesy'' citing between scientists, not always objectively justified, has also been pointed out \cite{raeyb}. The list of psychological pressures over the authors in the direction of increasing the number of citations can be completed by the need of pleasing the referees (prior to the revision or even after their suggestion) and the editors.

Summarizing, there are a number of reasons to include a reference in a paper \cite{weinstock1971} and not all of them have the meaning of recognizing prior knowledge to support the novelty of the paper \cite{raeya}. But all the pressures that can be identified acting on these reasons become catalytic factors for an increase in the number of references included. Therefore, the growing tendency that our data present is not surprising at all.

Added to all factors previously discussed, the enormous increase in the document accessibility generated by the expansion of the Internet can also be part of the pressure towards the increase in citation rates. Nowadays, literature can be reviewed and accessed from the office or the laboratory through extremely easy interfaces. However, not so long ago, these tasks required going to the library and use much more unfriendly and inefficient tools . Therefore, it is very easy today to find a reference when writing a paper to support ad hoc any argument. This phenomenon took place in a relative short period of time around 2000. Thus, its effect in the referencing rate would be an increase in the citations per paper (or per page of a paper). The data here presented are consistent with this hypothesis, although cannot be considered conclusive. Figure~4 shows a clear inflexion around 2000, but other similar inflexions are present in the studied period. Data on the number of pages contain much more errors than those of the references; this generates a noise superimposed to the general tendencies in Figure~4 that makes difficult to obtain positive conclusions. Much clearer is the year 2000 effect observed in Figure~1, when considering data of references per paper without pages correction. In some of the journals (see Figure~2) the slope change around 2000 is particularly strong, but variations among journals are significant, making difficult to establish a definitive conclusion. In any case, although not definitively, the year 2000 effect is strongly supported by our data.

Will this growth trend of the number of references continue indefinitely? When Yitzhaki and Ben-Tamar asked themselves this question in 1991 \cite{raeyb} they stated:

\begin{quote}
It is difficult to foretell, but one may nevertheless anticipate that at some point in time the growth trend will eventually slow down, and enter a ``saturation phase'', in which the number of references per paper will grow much slower each year.
\end{quote}

\noindent More than 10 years after, we see that the ``saturation phase'' is far from taking place. On the contrary, the citation rate increase keeps accelerating. The technical impossibility of an indefinite growing leads to propose a saturation phase sometime in the future. However, many pressures can be found towards citation increase and none in the opposite direction, making difficult to foresee citation rate stagnation. It is time to stop and reflect: where will we end up? Is this growth necessary or even reasonable? Is the means becoming an end?

The scientific communication standards, often mocked as the ``publish or perish system'', may suffer significant changes in the near future due to digital repositories, open access, social networks and the like. In any case, it is easy to predict that the reference to previous work will remain relevant, as it has been, at least, since the medieval tradition of stating the contributions as side notes to Aristotle work.

Finally, we have noted that the number of references included in this paper is slightly below our own discipline's average. We apologize for the inconvenience that this fact could cause to future researchers in the field.

\section{Conclusions}

This work analyzes the referencing patterns, from 1972 to 2013, in papers belonging to eight leading and long-standing engineering journals. The data retrieval methodology is based on a novel script able to automatically extract thousands of papers from the ISI WoK (easily adaptable to Scopus or other databases) in a few minutes. These are the main findings of our analysis:

\begin{itemize}

\item Along the same line as older studies in other disciplines, this paper has found a strong growth trend in the number of references made by engineering papers. In the period 1972-2000, the median has been doubled, reaching 16 references per paper. Not surprisingly, all the discussed factors affecting the referencing habits tend to push up these numbers.

\item Although other studies predict an eventual slowdown in the growth trend, there is still no sign of the so called \textit{saturation phase}. Quite the contrary, it has been found a growth acceleration after the year 2000, with a median that reaches 25 references per paper at present. Among all factors, the Internet, which leads to a greater access to information, stands as the most probable source of this effect.

\item Moreover, it has been demonstrated that the aforementioned effects exist independently of the parallel growth in the number of pages per paper.

\end{itemize}

\noindent According to prior studies, it would not be farfetched to think that these findings are valid for other fields. Further research would be needed to ascertain this point.


\bibliographystyle{plain}
\bibliography{abainv}

\end{document}